\documentclass{aa}
\usepackage[pdftex]{graphicx}
\usepackage{natbib,amssymb,amsmath,url}
\usepackage{hyperref}
\usepackage[utf8]{inputenc}

\newcommand{\ie}{{\it i.e.}}
\newcommand{\eg}{{\it e.g.}}
\newcommand{\be}{\begin{equation}}
\newcommand{\ee}{\end{equation}}

\begin{document}

\title{Constraints on primordial black hole dark matter from Galactic center X-ray observations}
\titlerunning{Constraints on PBH DM from GC X-rays}

\author{Andi Hektor \inst{1}, Gert H\"utsi \inst{2,1}, Martti Raidal \inst{1}}
\institute{National Institute of Chemical Physics and Biophysics, R\"avala 10, 10143 Tallinn, Estonia 
\and Tartu Observatory, University of Tartu, T\~oravere, 61602 Tartumaa, Estonia
\\ \email{andi.hektor@cern.ch}\\ \email{gert.hutsi@to.ee} \\ \email{martti.raidal@cern.ch} 
}
\date{Received / Accepted}

\abstract
{Surprisingly high masses of the black holes inferred from the Laser Interferometer Gravitational-Wave Observatory (LIGO) and Virgo gravitational wave measurements have lead to speculations that the observed mergers might be due to ${\cal O}(10) M_\odot$ primordial black holes (PBHs). Furthermore, it has been suggested that the whole amount of dark matter (DM) might be in that exotic form.}
{We investigate constraints on the PBH DM using NuSTAR Galactic center (GC) X-ray data.}
{We used a robust Monte Carlo approach in conjunction with a radiatively inefficient PBH accretion model with commonly accepted model parameters. Compared to previous studies we allowed for multiple forms of DM density profiles. Most importantly, our study includes treatment of the gas turbulence, which significantly modifies the relative velocity between PBHs and gas.}
{We show that inclusion of the effects of gas turbulence and the uncertainties related to the DM density profile reduces significantly the gas accretion onto PBHs compared to the claimed values in previous papers. It is highly improbable to obtain accreting PBHs brighter than the NuSTAR point source limit using observationally
determined gas velocities.}
{One can safely conclude that GC X-ray observations cannot rule out ${\cal O}(10) M_\odot$ PBH DM.}
\maketitle

\section{Introduction}
The first direct detection of gravitational waves by LIGO~\footnote{www.ligo.org} collaboration and several subsequent detections by LIGO and Virgo~\footnote{www.virgo-gw.eu}, have revolutionized observational astronomy by opening a new window into the cosmos. Thus far five binary black hole (BH) and a single binary neutron star mergers have been observed~\citep{2016PhRvL.116x1103A, 2016PhRvL.116f1102A, 2017ApJ...851L..35A, 2017PhRvL.119n1101A, 2017PhRvL.119p1101A}. Although the measurements have been fully consistent with general relativity, and thus have severely constrained modified gravity models, the large inferred BH masses have been somewhat surprising.~\footnote{Considering the complexity of star formation and binary stellar evolution, and the fact that this is the first time we are able to  probe BH-BH systems directly, one should not immediately draw too strong conclusions; see, \eg,~\citet{2016Natur.534..512B, 2017ApJ...851L..25F, 2017arXiv170909197D, 2018arXiv180103099L}.}

This has initiated vigorous speculations that the observed mergers might be between ${\cal O}(10) M_\odot$ primordial BHs (PBHs)~\citep{Hawking:1971ei, Carr:1974nx, Carr:1975qj, 1975A&A....38....5M, CHAPLINE:1975aa} and, furthermore, that the whole dark matter (DM) might be in the form of PBHs in that mass range~\citep{Kashlinsky:2016sdv,Bird:2016dcv,Clesse:2016vqa,Sasaki:2016jop}  (for a review see, \eg, \citet{Sasaki:2018dmp}). However, several recent works have shown that this interpretation is strongly constrained. In the case of DM in the form of ${\cal O}(10) M_\odot$ PBHs, the increased two-body interaction rate has potentially significant impact on dynamics of the stellar systems~\citep{2016ApJ...824L..31B, 2017PhRvL.119d1102K}. Also the gravitational lensing gets modified~\citep{2017ApJ...836L..18M, 2017arXiv170610281D, 2018PhRvD..97b3518O, 2017arXiv171202240Z}. The present constraints on PBH abundances from various lensing experiments and dynamical observations  are re-derived in~\citep{Carr:2017jsz} for various non-monochromatic PBH mass functions.  
Those constraints must be supplemented with bounds arising from  non-observation of the gravitational wave background~\citep{Raidal:2017mfl} and from the measured LIGO rate~\citep{2017PhRvD..96l3523A, 2018arXiv180505912A}. 

Thus the accumulated experimental data collectively constrain the fraction of PBH DM, $f_{\rm PBH}$, to be below unity, barring scenarios in which the radiation induced by PBH is strongly modified~\citep{2018arXiv180207728R} or the results of lensing experiments are misinterpreted~\citep{2018PDU....19..144G, 2017arXiv171110458C, 2017arXiv171206574G}.

The aim of this work is to revisit the constraints on PBH abundance arising from the photon flux created by accreting PBHs in our Galaxy. Our approach, based on the Galactic measurements, is complementary to those previously adopted in the literature to constrain the PBH abundance from the cosmic microwave background observations~\citep{Ricotti:2007au, 2016arXiv161207264H, Ali-Haimoud:2016mbv,Poulin:2017bwe} and from recent global 21 cm measurements~\citep{Hektor:2018qqw}. The authors of~\citet{2017PhRvL.118x1101G} have claimed very strong constraints on ${\cal O}(10) M_\odot$ PBH DM using X-ray and radio observations of the Galactic center (GC). In this paper we reconsider the GC X-ray constraints on ${\cal O}(10) M_\odot$ PBH DM taking into account physical effects that were overlooked in the latter work. Compared to~\citet{2017PhRvL.118x1101G} we include in our analysis turbulent gas motions inside molecular clouds and also allow for uncertainties in the DM density profile of our Galactic halo. We show that inclusion of those ingredients, in particular the measured velocities of the gas motion,
strongly suppress the PBH accretion and remove the previously claimed strong bounds arising from the NuSTAR GC X-ray data. 

Bounds on PBHs based on accretion arguments using X-ray data have also been found in~\citet{2017JCAP...10..034I}. There the authors use the extragalactic luminosity function of X-ray binaries as determined by~\citet{2012MNRAS.419.2095M} and demand that the accreting PBH population should not overshine this astrophysical component.

The paper is structured as follows. In Section~\ref{sec2} we briefly sketch our physical assumptions and calculation method, our main results are presented in Section~\ref{sec3}, and the discussion and summary are in Section~\ref{sec4}.

\section{Model details and parameter assumptions}\label{sec2}
Black holes are only visible in electromagnetic (EM) radiation if they accrete  a sufficient amount of baryonic material. The only EM-visible stellar mass BHs we know are in binary systems, where the companion star can provide sufficient mass transfer. However, owing to the relative diluteness of ${\cal O}(10) M_\odot$ PBH DM and thus rather negligible two-body scattering with stars in Milky Way (MW) size galaxies, it is usually not an option for PBHs to get incorporated into binary systems with ordinary star as a companion. Thus PBHs have to accrete matter directly from the interstellar medium (ISM).

In this case it is relevant to apply the Bondi accretion model~\citep{Bondi:1952ni} as a useful starting point. However, the Bondi mass accretion rate $\dot M_B$ has to be reduced by a factor $\lambda$ to be consistent with the non-observation of significant population of isolated neutron stars~\footnote{We note that in the context of the accretion flow described by the ADAF model, it is natural to try to use a population of isolated neutron stars to gain access to the possible values for the $\lambda$ parameter. In particular, neutron stars are almost as compact as BHs. These stars have, for their limited range of masses, radii that are only a factor of $\sim 3$ larger than those of the would-be BHs with similar masses (\ie, comparable to the size of the last stable orbit of the nonrotating BH). Compared to BHs, however they have a surface that lights up, and thus provides a direct way to probe the inflow of advected material.}. The present upper bound on $\lambda$ is $\sim 10^{-3} - 10^{-2}$~\citep{2003ApJ...594..936P}. To obtain model for the emitted EM radiation, the description of the mass accretion rate has to be augmented with a description of the radiative efficiency. As common in cases with low mass accretion rate and small opacity, we assume an advection dominated accretion flow (ADAF) model~\citep{Narayan:1994xi}. The radiative efficiency $\eta$ (bolometric luminosity $L_{\rm bol}=\eta \dot{M}c^2$) is often approximated as
\be\label{eq1}
\eta = 0.1 \times
\begin{cases}
\frac{\dot{m}}{\dot{m}_{\rm crit}} & \text{if $\dot{m} \le \dot{m}_{\rm crit}$} \\
1 & \text{if $\dot{m} > \dot{m}_{\rm crit}$}\,,
\end{cases}
\ee
where $\dot{m}$ is the mass accretion rate in units of Eddington rate, \ie, $\dot{m}\equiv \dot{M}/\dot{M}_{\rm Edd}$, and $\dot{m}_{\rm crit}\simeq 0.01$ \citep[\eg,][]{Narayan:2008bv}.
Taking $\dot{M}=\lambda \dot{M}_B$ and assuming the above radiative efficiency we write the accreting PBH luminosity as~\footnote{We have assumed that we are always in a regime where the mass accretion rate is below $1\%$ of the Eddington's and thus radiative efficiency is given by the first line of Eq.~(\ref{eq1}).}
\be\label{eq2}
L\simeq 8.8\times 10^{29}\frac{\rm erg}{\rm s} f \lambda^2 
\left(\frac{M_{\rm BH}}{10\,M_\odot}\right)^3
\left(\frac{n_{\rm H}}{1\,{\rm cm}^{-3}} \right)^2
\left(\frac{v_{\rm eff}}{10\,{\rm km\,s^{-1}}}\right)^{-6},
\ee
where $M_{\rm BH}$ is a BH mass, $f$ is a fraction of energy going to a selected energy band, $n_{\rm H}$ is the hydrogen number density, and $v_{\rm eff}$ is the {\it total} relative velocity of a BH with respect to the gas. This can be modeled to contain various components: (i) motion of the BHs, (ii) thermal motion of the gas,  and (iii) turbulent motion of the gas. We also denote a dangerous divergence due to $L \propto v^{-6}$, which needs careful treatment in any realistic physical system.

The highest chance of observing a bright PBH is toward a dense molecular region of the GC, or the so-called central molecular zone (CMZ), where conditions for both the high number density of PBHs and a dense surrounding medium are satisfied. Because of the measured coldness of the dense medium, we can safely neglect thermal motions of the gas to estimate $v_{\rm eff}$ in Eq.~(\ref{eq2}). However, turbulent gas motions in CMZ are far from negligible. For example, the 1D velocity dispersions of $2.6$-$53$~km/s with the median value of $9.8$~km/s have been inferred in~\citet{2016MNRAS.457.2675H}. Under these assumptions $v_{\rm eff}$ in Eq.~(\ref{eq2}) can be approximated as $v_{\rm eff}\approx (v_{\rm BH}^2 + v_{\rm turb}^2)^{1/2}$. The BH velocity $v_{\rm BH}$ is assumed to follow the Maxwell-Boltzmann (MB) distribution, where the MB scale parameter (1D velocity dispersion) is obtained by solving spherically symmetrized Jeans equation~\citep[\eg,][]{1987gady.book.....B} with two components: the baryonic bulge and the DM halo.

We assume an isotropic velocity distribution, \ie, $\sigma_r=\sigma_{\theta}=\sigma_{\phi}$. The PBH velocity dispersion is obtained by solving
\be\label{eq3}
\sigma_r^2(r)=\frac{1}{n_{\rm PBH}(r)}\int\limits_r^{\infty}n_{\rm PBH}(x)\frac{GM_{\rm tot}(<x)}{x^2}{\rm d}x\,,
\ee 
where the tracer density is that of the PBHs, $n_{\rm PBH}\,(\propto\rho_{\rm DM})$, but the mass inside radius $r$ should contain all the assumed components, $M_{\rm tot}(<r)=M_{\rm DM}(<r)+M_{\rm bulge}(<r)$.

To be more precise, a subdominant contribution from the baryonic disk should also be included. For simplicity, we neglect it in our study, which makes our values for $v_{\rm BH}$ somewhat underestimated and thus more conservative. The density profile of the DM halo is allowed to have two analytic forms: Navarro-Frenk-White (NFW)~\citep{1997ApJ...490..493N} and Einasto~\citep{1965TrAlm...5...87E}. For the GC gas distribution we use an analytic fitting form from~\citet{2007A&A...467..611F}. There the analytic spatial model for the mean gas densities is provided, but ISM is supposed to have a broad hierarchy of densities. 
In the case where $v_{\rm turb}=0$, it turns out that the Bondi accretion in conjunction with the low velocity tail of the MB distribution has a particular scaling property: the number of bright accreting PBHs above a fixed luminosity is independent of the hierarchy.~\footnote{Assuming independence of the velocity and gas density distributions ($f_{\rm MB}(v)$ and $f(n_{\rm H})$, respectively) and that the PBH luminosity follows a deterministic relation given by Eq.~(\ref{eq2}), we can write the joint probability density function as $\mathcal{F}(L,n_{\rm H},v)=f_{\rm MB}(v)f(n_{\rm H})\delta(L-\kappa n_{\rm H}^2v^{-6})$, where $\kappa$ is a constant factor and $\delta$ is the Dirac delta function. By integrating over $n_{\rm H}$ we obtain the joint luminosity and velocity distribution $\mathcal{F}(L,v)$ from which the luminosity probability function can be obtained as as $f(L)=\int \mathcal{F}(L,v)\,{\rm d}v=\int f_{\rm MB}(v)f(v^3\sqrt{L/\kappa})\,{\rm d}v$. To a good approximation $f_{\rm MB}(v)$ is given as $f_{\rm MB}(v)=\sqrt{2/\pi}\cdot v^2/\sigma^3$, since even for the very dense molecular cloud cores with $n_{\rm H}\sim 10^5$~cm$^{-3}$ the threshold velocity below which the luminosity exceeds the NuSTAR limit is $v\sim 35$~km/s, \ie, comfortably smaller than the 1D velocity dispersion shown in Fig.~\ref{fig:velocity_disp}. As such, it is easy to see that $f(L)$ indeed does not depend on gas density distribution, $f(L)=\frac{1}{\sigma^3}\sqrt{\frac{2}{\pi}}\int v^2 f(v^3\sqrt{L/\kappa})\,{\rm d}v=\frac{1}{3\sigma^3}\sqrt{\frac{2\kappa}{\pi L}}$. The number of bright objects in volume $\Delta V$ follows as $N=n_{\rm PBH}\Delta V f(> L_{\rm NuSTAR})$, \ie, indeed independent of the gas number density distribution function.} However, if $v_{\rm turb}> 0$ this scaling property gets broken and for more reliable treatment there should be a model for the small-scale gas density distribution. We performed our model calculations for two separate cases where the small-scale gas density distribution is assumed, first to  follow a uniform distribution within a resolution element with the mean density given by the analytic model of~\citet{2007A&A...467..611F}, and second has a power-law probability distribution function $f(n_{\rm H})\propto n_{\rm H}^{-\beta}$ with a mean given by the model of~\citet{2007A&A...467..611F}. We take $\beta=2.8$, which is the value typical for the giant molecular clouds, \eg,~\citet{1999ptep.proc...61B,2002MNRAS.334..553A}.

\section{Main results}\label{sec3}

It is well known that because of the low level of contaminating backgrounds and absorption, a search for accreting compact objects is especially efficient in the X-ray band of the EM spectrum. As such we compare our PBH DM model predictions with the NuSTAR~\footnote{www.nustar.caltech.edu} GC survey, which has lead to a detection of $\sim 70$ X-ray point sources. The NuSTAR GC survey has point source limits $4\times 10^{33}$ and $8\times 10^{33}$ erg/s in the $3$-$10$~keV and $10$-$40$~keV bands, respectively~\citep{2016ApJ...825..132H}. Alternatively, it would be possible to use Chandra observations of the GC~\citep{2009ApJS..181..110M}, which has approximately an order of magnitude higher sensitivity in softer X-ray band $0.5$-$8$~keV. However, in softer bands, the number of X-ray sources is steeply rising (mostly due to cataclysmic variables), and thus it is significantly easier to hide a subdominant accreting PBH population. Thus in the following we present our results based on the NuSTAR observations.

\begin{figure}
  \includegraphics[width=0.5\textwidth]{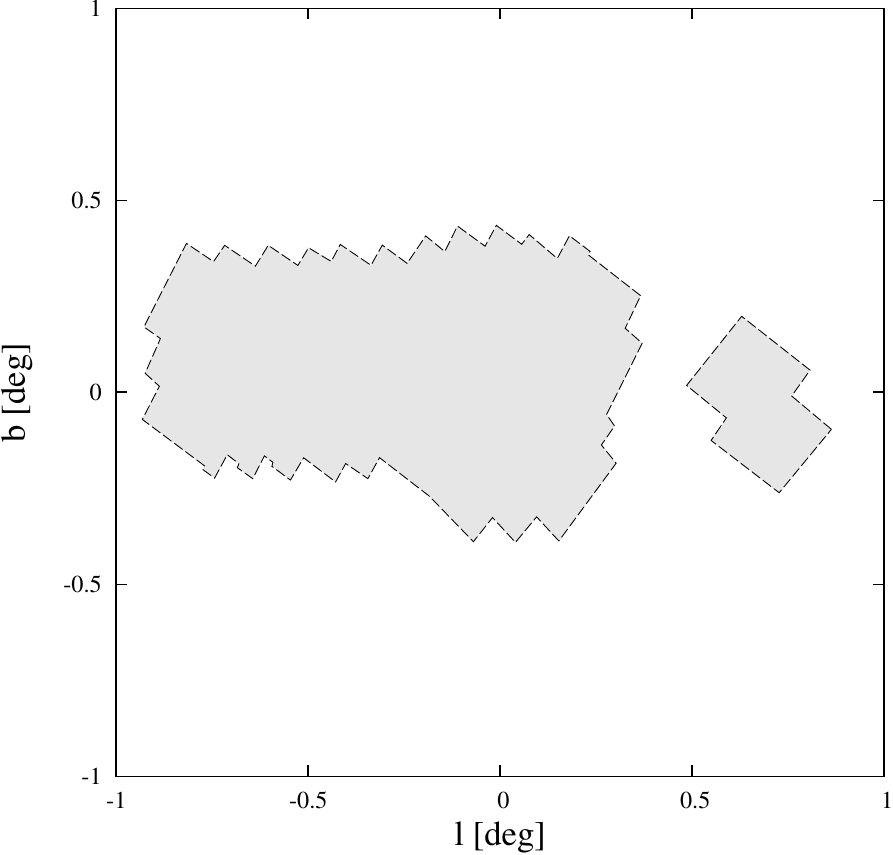}
  \caption{Approximate spatial mask of the NuSTAR GC survey shown in Galactic coordinates $(l,b)$.}
  \label{fig:NuSTAR-mask}
\end{figure}

\begin{figure}
  \centering
  \includegraphics[width=0.5\textwidth]{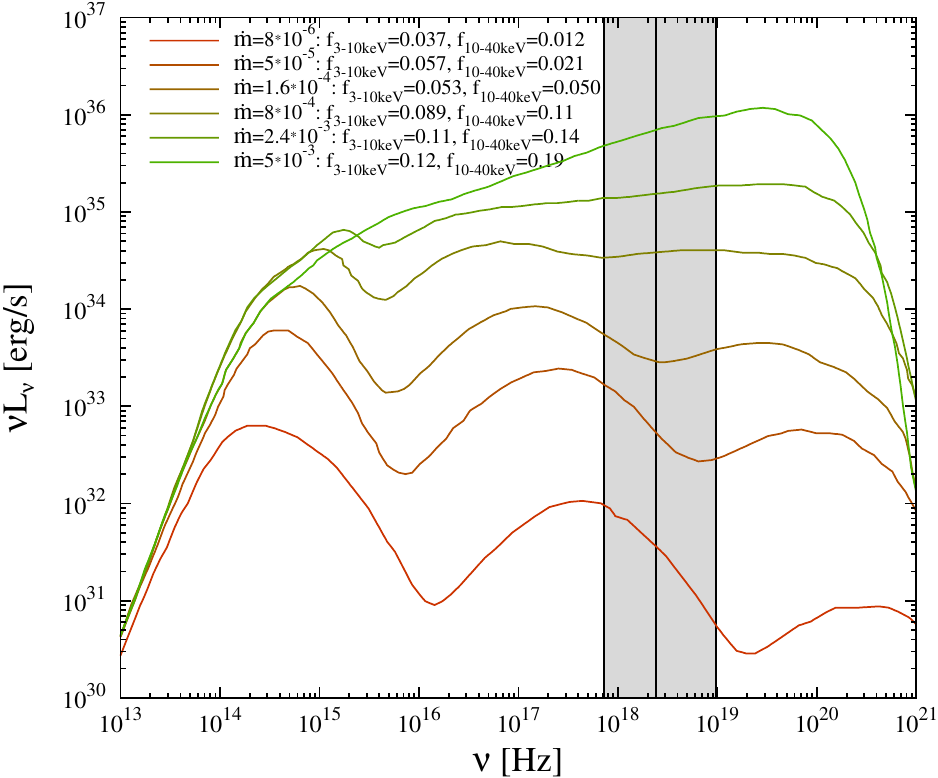}
  \caption{Spectra from ADAF model for various values of the specific accretion rate $\dot{m}\equiv \dot{M}/\dot{M}_{\rm Edd}$~\citep{2014ARA&A..52..529Y}. The parameters $f$ denote the fractions of energy going to the soft and hard NuSTAR bands shown with the gray-shaded regions.}
  \label{fig:spectra}
\end{figure}

Fig.~\ref{fig:NuSTAR-mask} shows an approximate mask we used for the NuSTAR GC survey derived from the results presented in~\citep{2016ApJ...825..132H}).
In Fig.~\ref{fig:spectra} we show the ADAF model spectra ($M_{\rm BH}=10M_\odot$) for different values of the specific accretion rate $\dot{m}\equiv \dot{M}/\dot{M}_{\rm Edd}$ taken from~\citet{2014ARA&A..52..529Y}; see Fig. 1 therein. The fractions of energy going to the soft and hard NuSTAR bands (see the gray shaded regions) are also shown for each of the model curves. We note that the ADAF model depends on various input parameters such as (i) viscosity parameter, (ii) magnetization parameter, (iii) electron heating parameter, and (iv) wind parameter, which in this particular case have values $\alpha=0.1$, $\beta=9$, $\delta=0.5$, and $s=0.4$, respectively. In case a jet with a nonthermal population of electrons appears or if the hot flow itself contains a nonthermal component, the prominent inverse Compton bumps of Fig.~\ref{fig:spectra} get significantly smoothed out. For example, such a nonthermal electron component is required for the ADAF model to provide a satisfactory fit to the available SgrA* data~\citep{2003ApJ...598..301Y}.

\begin{figure}
  \includegraphics[width=0.48\textwidth]{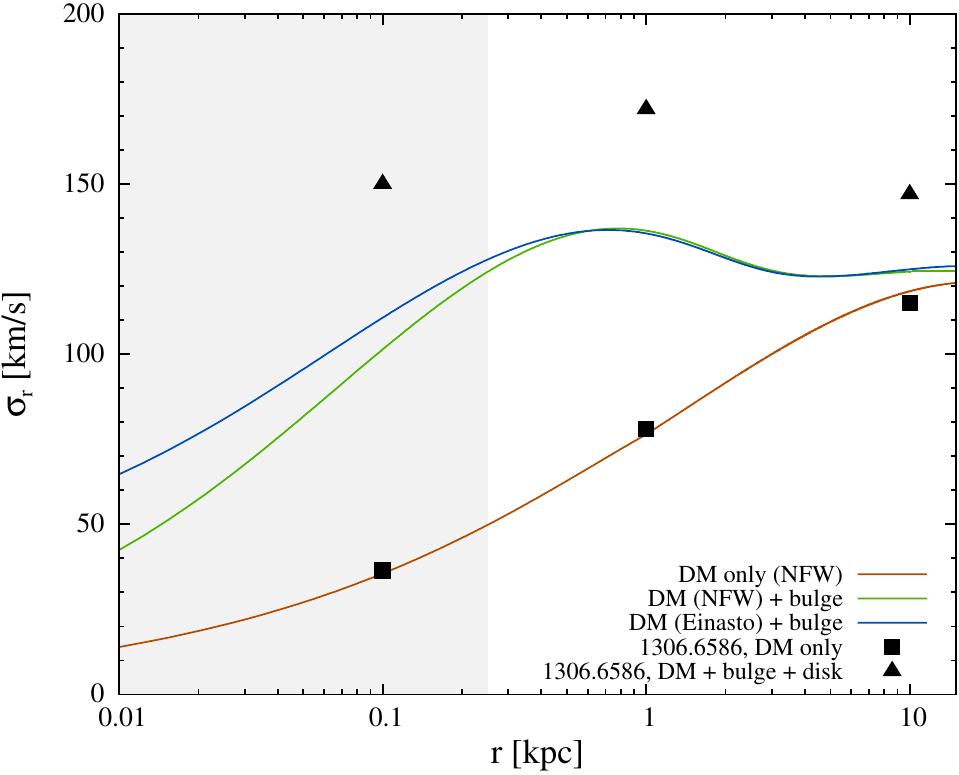}
  \caption{Velocity dispersion of the PBH DM. The solid curves show our estimate from the Jeans equation. Three cases are shown: the Galaxy consisting only of (i) NFW DM halo, (ii) NFW DM halo + baryonic bulge, and (iii) Einasto DM halo + baryonic bulge. The triangles and squares denote the DM velocity dispersions taken from~\citet{2013JCAP...09..005F}. Neglecting the baryonic disk leads to a $\sim 25\%$ underestimation of the DM velocity dispersion, \ie, a factor of $\sim 3$ overestimation of the accreting PBH luminosity. Replacing NFW with the Einasto density profile leads to a factor of $\sim 1.5$ increase in the velocity dispersion in the very central parts. The light gray shaded region shows an approximate extent of the CMZ.}
  \label{fig:velocity_disp}
\end{figure}

Fig.~\ref{fig:velocity_disp} shows the PBH DM velocity dispersions derived from spherically symmetric Jeans equation. The three solid lines show the Galactic model consisting only of (i) NFW DM halo, (ii) NFW DM halo + baryonic bulge, and (iii) Einasto DM halo + baryonic bulge. The profile parameters relevant for the Galactic DM halo are taken from~\citet{2011JCAP...03..051C}. For the baryonic bulge an analytic density distribution from~\citet{2011MNRAS.414.2446M} is assumed, while for simplicity spherically symmetric approximation is taken by replacing the scale radius $r_0$ with the following geometric mean $\sqrt[3]{r_0^2\times qr_0}$ ; the Galactic bulge has an approximate aspect ratio of $q\simeq1/2$. The value $r_{\rm cut}$ is also replaced similarly.
For comparison, the triangles and squares in Fig.~\ref{fig:velocity_disp} denote the DM velocity dispersions taken from~\citet{2013JCAP...09..005F}, in which more detailed modeling is performed. Compared to~\citet{2013JCAP...09..005F} we assumed spherical symmetry and completely ignored contribution from the baryonic disk, which can be seen to lead to $\sim 25\%$ underestimation of the DM velocity dispersion, \ie, a factor of $\sim 3$ overestimation of the accreting PBH luminosity. Also,  replacement of NFW with the Einasto halo leads to a factor of $\sim 1.5$ increase in the velocity dispersion in the very central region. However, at extremities of the CMZ the effect is far less pronounced, but it is important to keep in mind that, according to Eq.~(\ref{eq2}), velocity rises as $L \propto v^{-6}$; the approximate
size of the CMZ is shown as a light gray shaded region. 

\begin{figure*}
  \centering
  \includegraphics[width=\textwidth]{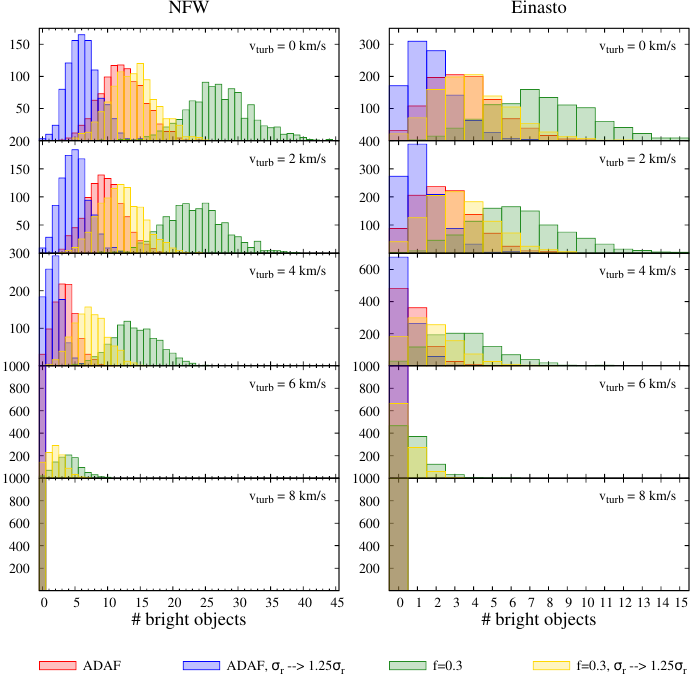}
  \caption{Distributions for the number of bright PBHs above the NuSTAR point source detection limit for different levels of turbulence in the CMZ ($0$-$5$~km/s) and for different DM profiles (NFW, Einasto). Each distribution is derived from 1000 independent Monte Carlo realizations. The orange histograms represent the model with only a bulge and DM halo. The green histograms correspond to the case in which the MB 1D velocity dispersion has been increased by $25\%$ to correct for the missing Galactic disk contribution (see Fig.~\ref{fig:velocity_disp}). The uniform gas distribution within resolution element is assumed.}\label{fig:MC}
\end{figure*}

\begin{figure*}
  \centering
  \includegraphics[width=\textwidth]{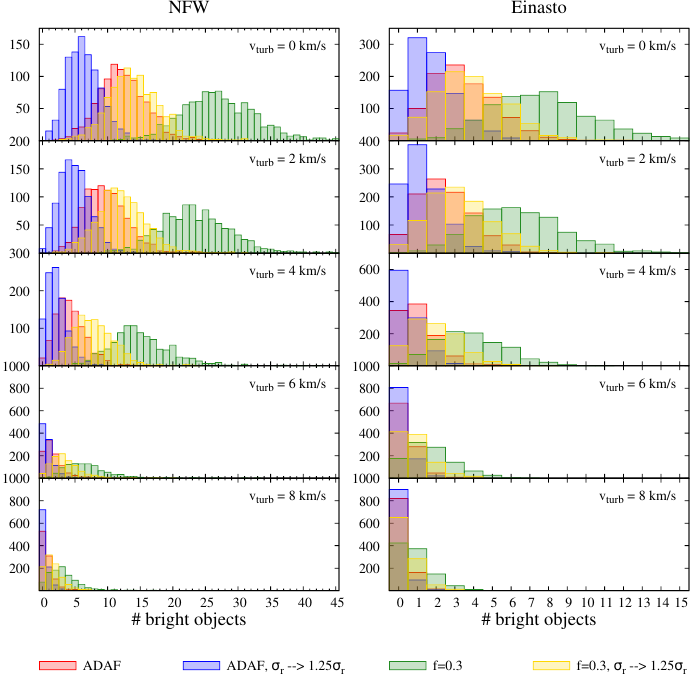}
  \caption{Analog of Fig.~\ref{fig:MC} for the model with a small-scale, power-law gas density distribution.}\label{fig:MC2}
\end{figure*}

Even though the velocity dispersions for the NFW and Einasto profiles shown in Fig.~\ref{fig:velocity_disp} at $r \sim 0.1$~kpc differ only by $\sim 10\%,$ it turns out that the more central regions, where the velocity dispersions and the DM density profiles differ more strongly, are responsible for very large differences we obtain between the NFW and Einasto cases (see Figs.~\ref{fig:MC} and~\ref{fig:MC2}).

Now we have all the ingredients available to proceed with our main calculations. We ran several Monte Carlo simulations, each containing 1000 independent realizations, the results of which are shown in Figs.~\ref{fig:MC} and~\ref{fig:MC2}. 
We assume that all the DM is in the form of PBHs. To speed up our computations we implemented a grid-based Monte Carlo by dividing the line-of-sight cone, which covers the full NuSTAR GC survey footprint into small cubic cells of size $\Delta V = 2\times 2\times 2$ pc$^3$. We looped over the cells and calculate the number of PBHs in each of these by drawing numbers from the Poisson distribution with mean $\overline{N}=\Delta V\times \rho_{\rm DM}/M_{\rm PBH}$, where $\rho_{\rm DM}$ is the assumed DM density and $M_{\rm PBH}$ the PBH mass. For each object drawn this way we generated their 3D velocity components by drawing numbers from the Gaussian with zero mean and dispersion taken from Eq.~(\ref{eq3}). By using a finer grid we confirm that the above cell size is sufficient for obtaining converging results.

In case the small-scale gas density distribution is taken to be a power law, $f(n_{\rm H})\propto n_{\rm H}^{-\beta}$, we draw the relevant number density as follows: $n_{\rm H}=n_{\rm H}^{0}(1-u)^{1/(1-\beta)}$ where $n_{\rm H}^{0}\equiv\frac{\beta-2}{\beta-1} <n_{\rm H}>$. The value $u$ is a random variable following the standard uniform distribution, \ie,~$u\sim U(0,1)$, $ <n_{\rm H}>$ is the mean gas density taken from~\citet{2007A&A...467..611F}; the power-law index is assumed to be $\beta=2.8$.

The results of our Monte Carlo calculations for the model with uniform gas density within the resolution element are shown in Fig.~\ref{fig:MC} in which the left- and right-hand panels correspond to the NFW and Einasto profiles, respectively. The red histograms represent the model with only the bulge and DM halo. The blue histograms correspond to the case in which the MB 1D velocity dispersion was increased by $25\%$ to correct for the missing disk contribution (see Fig.~\ref{fig:velocity_disp}). To facilitate a comparison with the results of~\citet{2017PhRvL.118x1101G} the analogous green and yellow histograms assume instead a fixed spectral factor of $f=0.3$. From top to bottom the effective velocity for turbulent gas motions is allowed to vary in the range $0-8$~km/s with $2$~km/s step size. In these calculations, PBH mass of $30 M_\odot$ is assumed. The numbers of bright objects should be compared with the NuSTAR point source count of $\sim$70. Only in the case of NFW halo plus negligible turbulent gas motions, is it possible to obtain (depending on a particular model) at most $\sim 10-30$ PBHs above the NuSTAR point source limit. More realistically, stochastic treatment for turbulent gas motions should also be included, \ie, the resulting effective distribution should be calculated as a weighted average over the distributions shown in Fig.~\ref{fig:MC}. In case turbulent gas motions are modeled with MB distribution with a realistic dispersion of $10$~km/s, the total weight for the distributions of Fig.~\ref{fig:MC} (velocity range $0-8$~km/s) is only $\sim 14\%$. As a result, in this case PBHs brighter than the NuSTAR limit are only rarely obtained, and it is safe to conclude, contrary to the claims made in~\citet{2017PhRvL.118x1101G}, that the GC X-ray observations cannot rule out ${\cal O}(10) M_\odot$ PBH DM.

The analogous results for the model with a power-law gas density distribution are shown in Fig.~\ref{fig:MC2}. In order to ease comparison we kept the axes scales identical to those used in Fig.~\ref{fig:MC}. As expected, the results in case of small $v_{\rm turb}$ are practically the same as obtained for the model with a constant small-scale gas density. For $v_{\rm turb}\gtrsim 5$~km/s the results start to differ more visibly, namely, we obtain somewhat larger number of bright objects. However, this increase is very moderate, and thus our previous conclusions remain unchanged.

The above modeling has completely ignored possible two-body relaxation effects due to the granularity of the DM distribution. According to~\citet{2017PhRvL.119d1102K}, in the case of dwarf galaxies these effects are not negligible. However, it turns out that the relevant relaxation time for the MW size halo inside the central region comparable to the size of the CMZ is significantly larger than the Hubble time. Thus, the two-body scattering effects can be safely neglected. The relaxation only becomes noticeable in the very central (few tens of parsecs) region. In this region PBHs, which are significantly more massive than typical stars, would start to migrate to the center. The corresponding density profile would steepen by trying to approach the isothermal $\rho \propto r^{-2}$. However this does not increase the chances of obtaining bright PBHs, since the central subsystem would heat up and the $v^{-6}$ scaling of Eq.~(\ref{eq2}) would dominate above the rise from the steepening of the density profile.

\section{Discussion and summary}\label{sec4}
In this paper we investigated how well the Galactic X-ray measurements are able to constrain ${\cal O}(10 M_\odot)$ PBH DM. The probability of seeing bright accreting PBHs is highest for the GC, where the conditions for large PBH number density and high gas density, are simultaneously met. As such, we used the GC data from the NuSTAR X-ray telescope in our analysis.

A similar study has previously been performed by~\citet{2017PhRvL.118x1101G}.
Compared to this work, rather than having a single fixed form for the DM density profile, we investigate how much the results change by allowing a reasonable level of profile variability. The other, arguably the most important difference, is that our model includes treatment for the ISM gas turbulence. In addition, our treatment for the spectral factor $f$ of Eq.~(\ref{eq2}) is more complete: we calculate $f$ self-consistently using the ADAF spectral templates taken from~\citet{2014ARA&A..52..529Y}, whereas~\citet{2017PhRvL.118x1101G} assume a fixed value $f=0.3$.

In~\citet{2017PhRvL.118x1101G} ${\cal O}(10 M_\odot)$ PBH DM is claimed to be ruled out by up to $40\sigma$ using the GC X-ray data, \ie, this would correspond to around $\sim$1500 X-ray visible PBHs. In comparison, only in our most optimistic case with negligible gas turbulence and the NFW density profile, we obtain on average $\sim 10-15$ X-ray visible PBHs above the NuSTAR point source limit. For almost all the other cases in which a reasonable level of gas turbulence is allowed, hardly any bright PBHs are found. Thus the extraordinary strong claims made in \citet{2017PhRvL.118x1101G} are artifacts of the unphysical $v^{-6}$ singularity in the PBH luminosity estimation arising from their assumption that the accreted gas is standing still. We demonstrate that the inclusion of measured gas velocities removes this singularity, and as a result refutes their claims regarding strong bounds on the PBH abundance.

We also note that the use of the experimental {\it upper bound} for the accretion efficiency parameter of Eq.~(\ref{eq2}), \ie, $\lambda\sim10^{-2}$, as done in~\citet{2017PhRvL.118x1101G} in order to rule out PBH DM, does not sound correct. For this purpose actually a {\it lower bound} is needed. The latter cannot be reliably estimated since the accretion physics in a highly turbulent and magnetized ISM has very large uncertainties.

In this paper we opted to use a specific ADAF accretion model taken from~\citet{2014ARA&A..52..529Y} to model the energy distribution of the emitted radiation. To compare our results directly with the results presented in~\citet{2017PhRvL.118x1101G} we also used a simple model in which $30\%$ of the emitted radiation is assumed to fall into NuSTAR X-ray energy band. We stress that beyond ADAFs, where most of the energy gained by the accreting gas is simply carried beyond the BH event horizon, there are many other possible models for the accretion from the relatively dilute ISM. Such models include adiabatic inflow-outflow solution (ADIOS) and convection-dominated accretion flow (CDAF) type solutions~\citep{1999MNRAS.303L...1B,2000ApJ...539..809Q}, which are inefficient because most of the accreting mass is simply driven out, and thus the mass actually gained by the BH is much smaller than the large-scale mass accretion scale as estimated by $\dot M_B$. Also depending on a particular model, the resulting energy distributions for the emitted radiation can vary substantially. It is not the topic of this paper to study all these possibilities. The main message we would like to convey is that under these very large model uncertainties, together with the absence of the relevant phenomenology for the isolated stellar-mass BHs accreting from the ISM, it is impossible to convincingly rule out a possible existence of the ${\cal O}(10) M_\odot$ PBH DM by accretion arguments alone. 
In general we agree that under the specific assumptions made in~\citet{2017PhRvL.118x1101G} it is possible to obtain a measurable population of X-ray sources above the NuSTAR sensitivity limit; however the above result is very different when allowing for different DM profiles and taking the significant level of turbulence of the gas in the CMZ into account.
To conclude, it is fair to say that with the current state of knowledge about the isolated stellar-mass BHs accreting from the ISM, the present GC X-ray observations cannot rule out ${\cal O}(10) M_\odot$ PBH DM.

\begin{acknowledgements}
This work was supported by the grants IUT23-6, IUT26-2, PUT808, and by EU through the ERDF CoE program grant TK133 and by the Estonian Research Council via the Mobilitas Plus grant MOBTT5. AH thanks the Horizon 2020 program as the project has received funding from the program under the Marie Sklodowska-Curie grant agreement No 661103.
\end{acknowledgements}

\bibliographystyle{aa}
\bibliography{references}

\end{document}